\documentclass[prl,twocolumn,showpacs,preprintnumbers,amsmath,amssymb,superscriptaddress]{revtex4}
\usepackage{amsmath,graphicx}
\usepackage{color}

\begin{document}

\title{Curvature-induced frustration in the $XY$ model on hyperbolic
surfaces}

\author{Seung Ki Baek}
\email{garuda@tp.umu.se}
\affiliation{Department of Physics, Ume{\aa} University, 901 87 Ume{\aa},
Sweden}
\author{Hiroyuki Shima}
\email{shima@eng.hokudai.ac.jp}
\affiliation{Department of Applied Physics, Graduate School of Engineering,
Hokkaido University, Sapporo 060-8628, Japan}
\author{Beom Jun Kim}
\email{beomjun@skku.edu}
\affiliation{BK21 Physics Research Division and Department of Energy
Science, Sungkyunkwan University, Suwon 440-746, Korea}
\affiliation{Department of Computational Biology, School of Computer Science
and Communication, Royal Institute of Technology, 100 44 Stockholm, Sweden}

\begin{abstract}
We study low-temperature properties of the $XY$ spin model
on a negatively curved surface.
Geometric curvature of the surface gives rise to frustration
in local spin configuration, which results in the formation of 
high-energy spin clusters scattered over the system.
Asymptotic behavior of the spin-glass susceptibility
suggests a zero-temperature glass transition,
which is attributed to multiple optimal configurations
of spin clusters due to nonzero surface curvature of the system.
It implies that a constant ferromagnetic spin interaction on a regular
lattice can exhibit glasslike behavior without possessing any disorder
if the lattice is put on top of a negatively curved space
such as a hyperbolic surface.

\end{abstract}

\maketitle

The minimum energy principle that a physical system obeys
may be broken by applying external conditions.
One such condition, which has profound consequences in
the ergodicity of the system,
is geometric confinement to a curved surface.
A typical example is an ensemble of classical electrons 
confined to a spherical surface~\cite{thomson}.
It exhibits so many quasistable configurations
that a unique ground state is hardly observed 
within a feasible time scale~\cite{travesset}.
These long-lived states result from the incompatibility between 
crystalline order and surface curvature.
In fact, partial disorders such as defects and dislocations
are necessary in order for
local crystalline order to propagate through a curved surface,
wherein various possible configurations of disorders
cause many frustrated states.
Such a curvature-induced frustration has also been observed
on more general geometries~\cite{travesset,hexemer-kohyama-giomi}.

Another important consequence of nonzero surface curvature
is a breakdown of an {\it orientational} order.
When interacting constituents on a curved surface
have orientational degrees of freedom,
they can no longer show perfect orientational order.
The loss of perfect orientational order is due to the noncommutative property
of parallel transport of vectors~\cite{frankel}.
On a curved surface,
parallel transport of a vector
along a closed loop does not maintain its direction
but yields a rotation after the round trip.
This makes it impossible for all vectors
to orient the same direction,
yielding multiple frustrated states at low temperatures
even if the system does not possess any disorder~\cite{tarjus}.
These facts imply a novel class of orientational glasses
free from any kind of disorder,
which is in contrast with ordinary spin glasses
dominated by certain disorder (see, e.g., Refs.~\cite{nonrand} and \cite{ekim} for
attempts at glass formation in absence of intrinsic disorder).
Understanding the nature of such curvature-induced glass transition, if it exists,
should be crucial from viewpoints of statistical physics
and soft material sciences.
Particularly in the latter field,
systems with curved or fluctuating geometries
are accessible to synthesize~\cite{kamien-hemmen},
although the interplay between the geometry and thermodynamic properties of 
allowed ordered states still remains to be explored~\cite{nieves-skacej-frank}.

There are two well-established approaches to analyze the orientational order
of a physical system:
discrete lattice simulations and continuum limit approximations.
The former approach is preferable to study actual evolutions of
orientational order.
On curved surfaces, however,
one cannot construct regular lattices with congruent polygons, in general.
Hence, the resulting lattice usually involves structural defects
as mentioned in the first paragraph.
These defects may give 
additional contributions to the allowed orientational configuration,
thus should be removed when we are to extract purely curvature effects.
This can be achieved
by employing a surface having constant negative Gaussian curvature
called a hyperbolic surface~\cite{coxeter}.
This curved surface enables to construct
a wide range of regular lattices
(called a hyperbolic lattice) on it,
serving as a platform to address the issue.

In this Rapid Communication, we consider effects of curvature-induced frustration
on the orientational order in
the $XY$ spin model defined on a hyperbolic lattice.
Monte Carlo (MC) simulations are used to evaluate the spin configurations
at low temperatures
revealing the formation of high-energy spin clusters distributed 
in the system.
We propose that these configurations
are multiply degenerate due to curvature-induced frustration,
which may lead to a zero-temperature glass transition as
supported by calculating the spin-glass susceptibility.

\begin{figure}[ttt]
\includegraphics[width=0.17\textwidth]{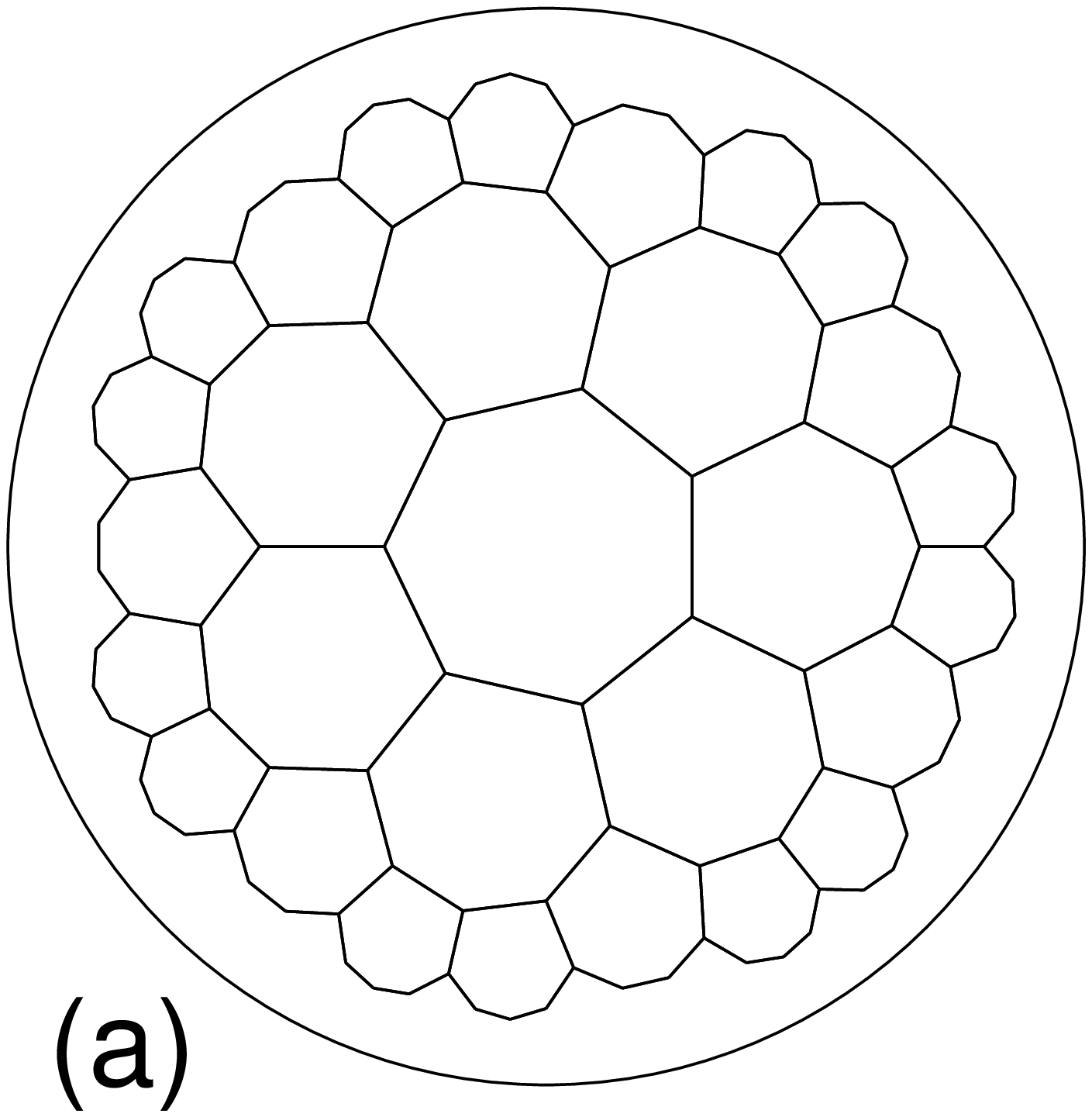}
\includegraphics[width=0.21\textwidth]{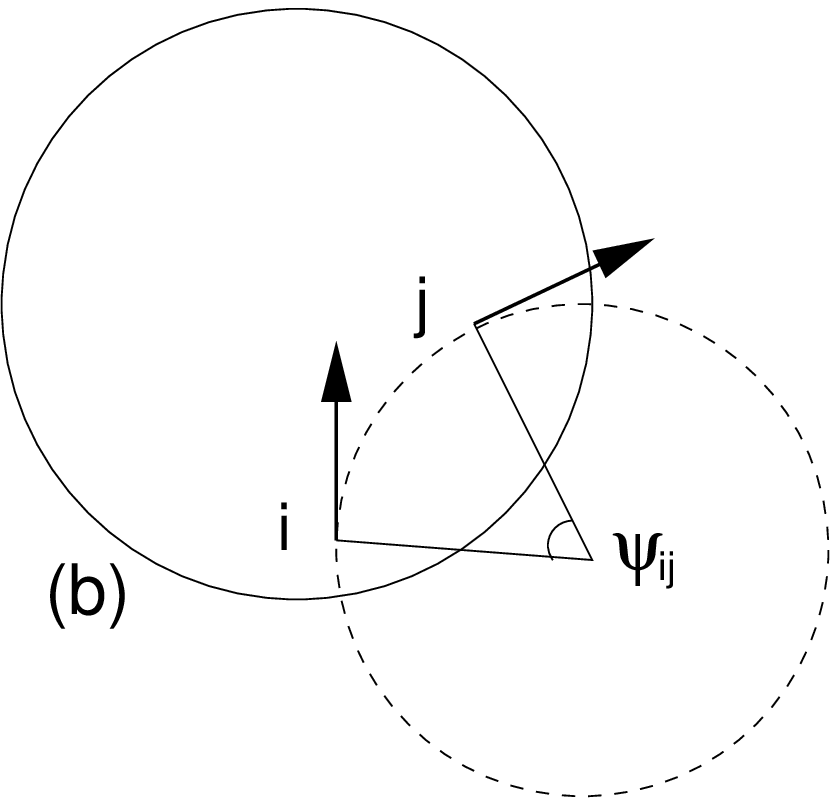}
\caption{(a) A heptagonal lattice with $l=3$ in the Poincar\'e disk. 
The circumference of the disk represents the points at infinity.
(b) Parallelism between two spins in the Poincar\'e disk.
Parallel transport of an arrow from $i$th to $j$th site along the geodesic 
curve (dotted) rotates the arrow by $\psi_{ij}$. 
}
\label{fig:hep}
\end{figure}

The hyperbolic lattice we have used is depicted in Fig.~\ref{fig:hep}(a)
in terms of the Poincar\'e disk representation~\cite{anderson}.
The lattice consists of equilateral heptagons in the metric of the hyperbolic surface
so that we call it a regular heptagonal lattice.
Seven vertices in the central heptagon make the first layer denoted by $l=1$,
which are surrounded in a concentric way
by the second $(l=2)$ and third $(l=3)$ layers.
The number of vertices $N(l)$ increases exponentially
with $l$, making this lattice infinite dimensional.
This infinite-dimensional property has been found to yield various
nontrivial thermodynamic properties of physical systems defined 
on this lattice~\cite{hyperb}.
The relevance to the mean-field behavior observed 
in small-world networks has also been discussed~\cite{hong-wsxy}.

Let us imagine two $XY$ spins $i$ and $j$ confined in the surface and a
geodesic between them.
On the analogy of the continuum limit, we suppose that
$i$'s phase is experienced by $j$ after parallel transported
to $j$'s position along the geodesic. 
The Hamiltonian is then given by
\begin{equation}
{\cal H} = -J \sum_{\langle ij \rangle} \cos(\phi_i - \phi_j - \psi_{ij}),
\label{eq:h2}
\end{equation}
where $J$ is the coupling constant, the sum is over all
nearest pairs in the system, and $\phi_i$ represents the spin variable at
the $i$th site.
Nonzero surface curvature manifests in the presence of the additional angle $\psi_{ij}$.
The angle $\psi_{ij}$ describes
the amount that $i$'s phase acquires after the parallel transport.
In the language of differential geometry, this additional angle
stems from the affine connection~\cite{frankel}
between two neighboring sites along the geodesic line.
Figure~\ref{fig:hep}(b) illustrates how to actually determine $\psi_{ij}$:
we obtain positions of $i$ and $j$ on the Poincar\'e disk by hyperbolic
tessellation~\cite{coxeter}.
If they lie on a line through the origin
of the Poincar\'e disk, the geodesic is represented by a straight line
yielding $\psi_{ij} = 0$. Otherwise, the geodesic appears as an arc of a
circle that meets the circumference of the Poincar\'e disk at a right
angle. The positions of $i$ and $j$ together with the origin of this circle
make an angle, which determines $\psi_{ij}$.

The spins at sites $i$ and $j$ are parallel
when $\phi_i - \phi_j - \psi_{ij} = 0$.
(Note that $\psi_{ij} = -\psi_{ji}$ to keep the spin-spin interaction
symmetric.)
Accordingly, parallel transport of the $i$th spin 
along seven edges of a heptagon alters the value of $\phi_i$
causing 
frustration in local spin configuration.
It should be emphasized that the local
frustration in our system has no relation
with any quenched disorder and originates purely from the intrinsic
geometry of the surface.
Such a disorder-free frustration is in contrast with disorder-driven
frustrations in random gauge $XY$ models~\cite{bjkimG1-jum-lhtang}.
It is also notable to see how this differs from such frustrations as
observed in the triangular antiferromagnetic Ising (AFI) model~\cite{ekim}:
while the hexagonal symmetry in AFI allows the presence of loose spins as
well as the fully-ordered configuration, none of these can be found in our
system.
We point out that the sum of $\psi_{ij}$ around each heptagon is conserved as
$\sum_P \psi_{ij} = 2 \pi f$ with $f=-1/6$
; here
$\sum_P$ represents the summation around a plaquette in a counterclockwise
direction, and $f$ characterizes the strength of frustration.
The quantitative invariance of $f$ for all constituent heptagons
is analogous to that in
the uniformly frustrated 
$XY$ model on a flat plane~\cite{nelson}.
The latter model describes a superconducting film penetrated by external
magnetic flux~\cite{teitel-teitel1983-choi1985}, 
and phase ordering in it is still under active investigations~\cite{hasen-bjkim}.
It is known that the uniformly frustrated $XY$ model exhibits glassy
behavior on a square lattice when $f$ is irrational~\cite{nelson,choi1985b-choi1987}.
We should notice, nevertheless, that
a direct analogy between the 
two systems
is hindered by topological differences between planar and hyperbolic
lattices. In fact, 
the hyperbolic lattices show glassy behavior
for a rational $f$ as demonstrated below.

\begin{figure}[ttt]
\includegraphics[width=0.17\textwidth]{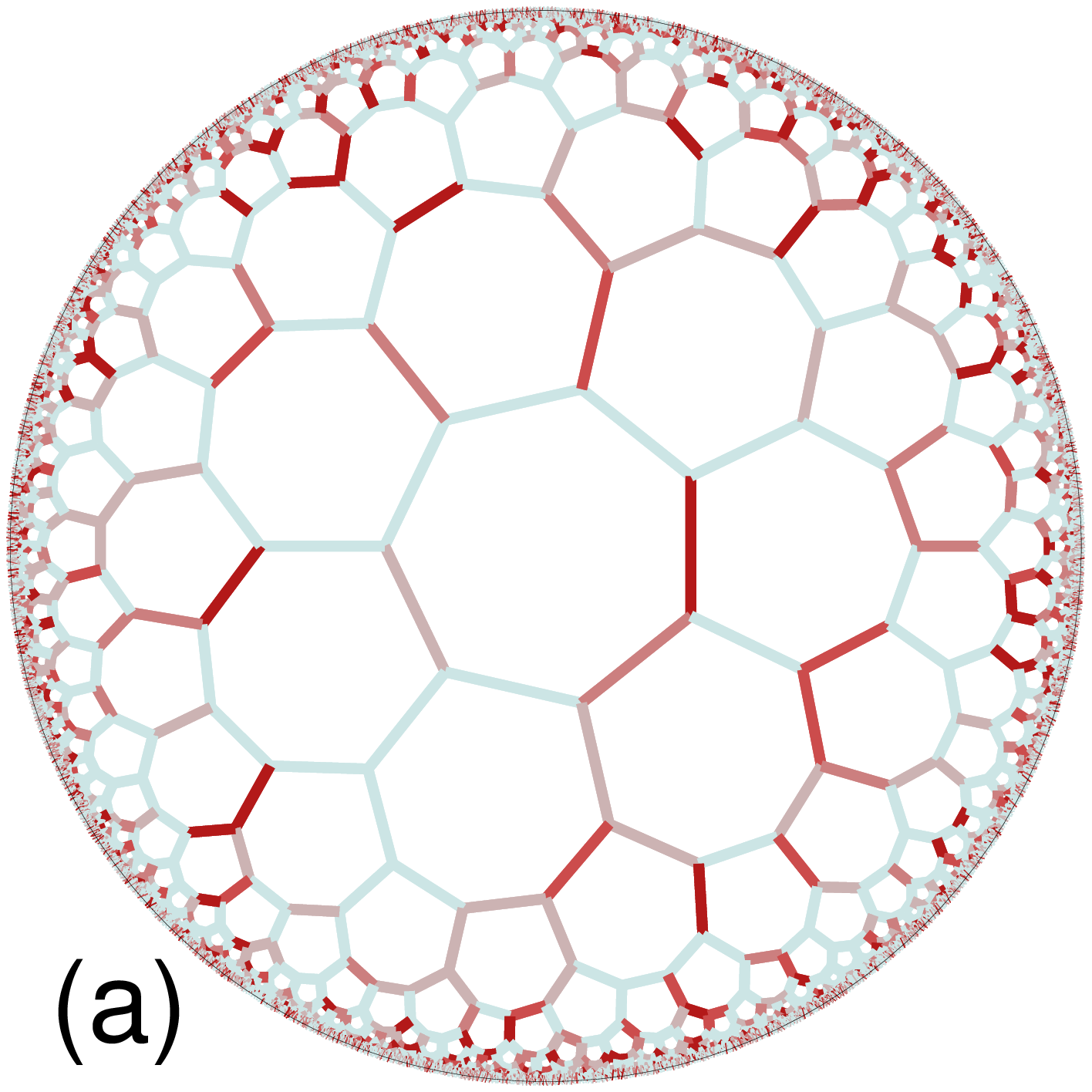}
\includegraphics[width=0.17\textwidth]{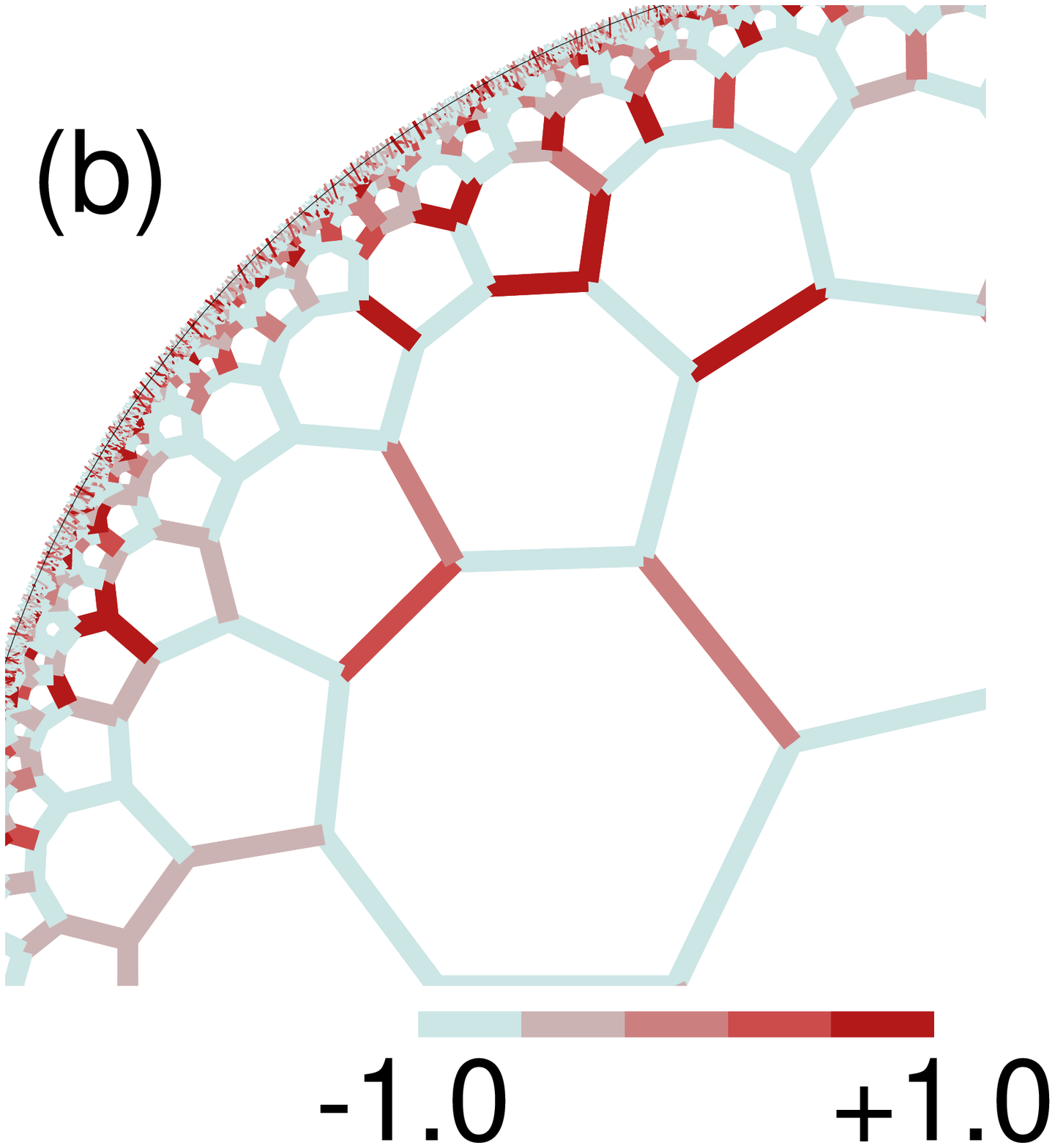}\\
\includegraphics[width=0.17\textwidth]{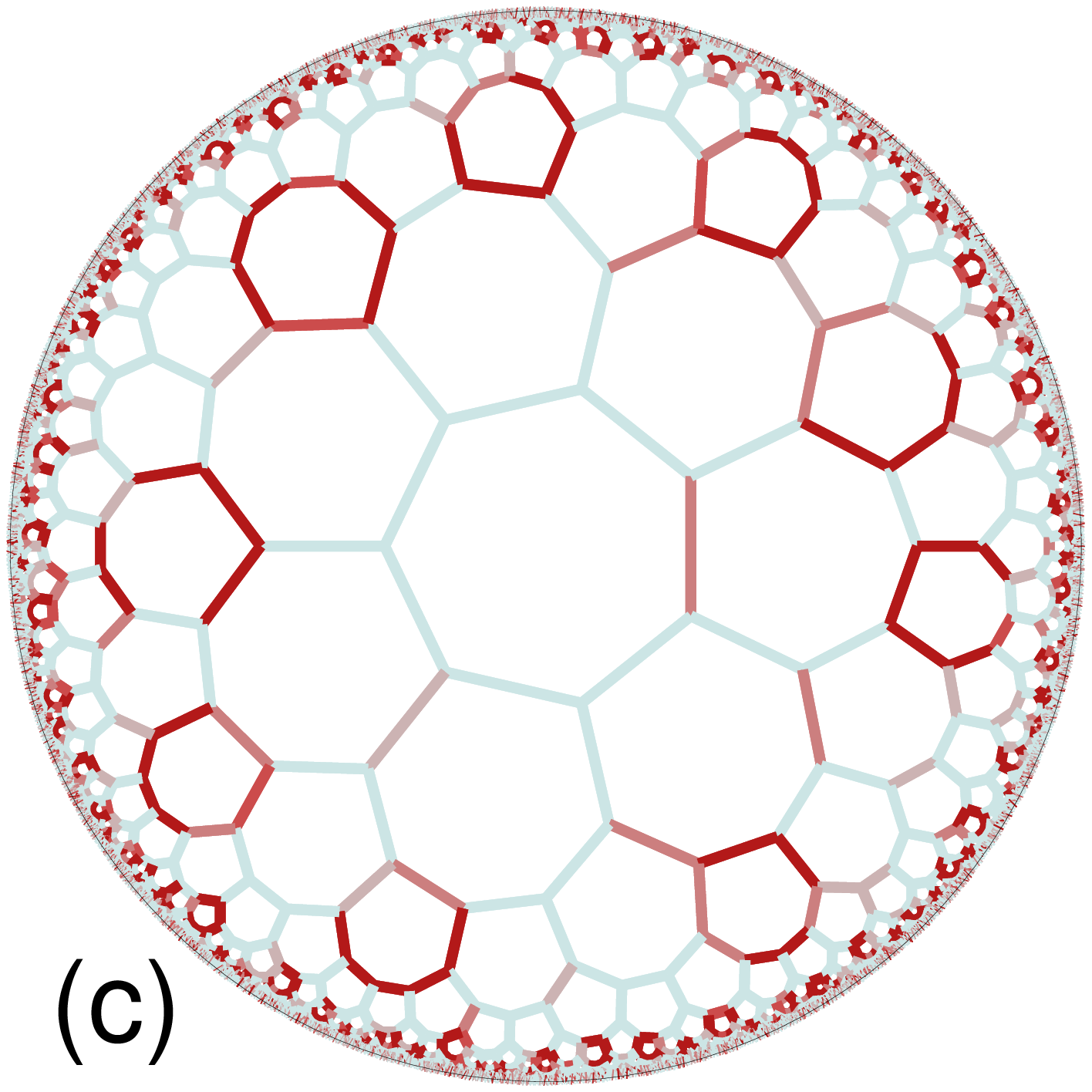}
\includegraphics[width=0.17\textwidth]{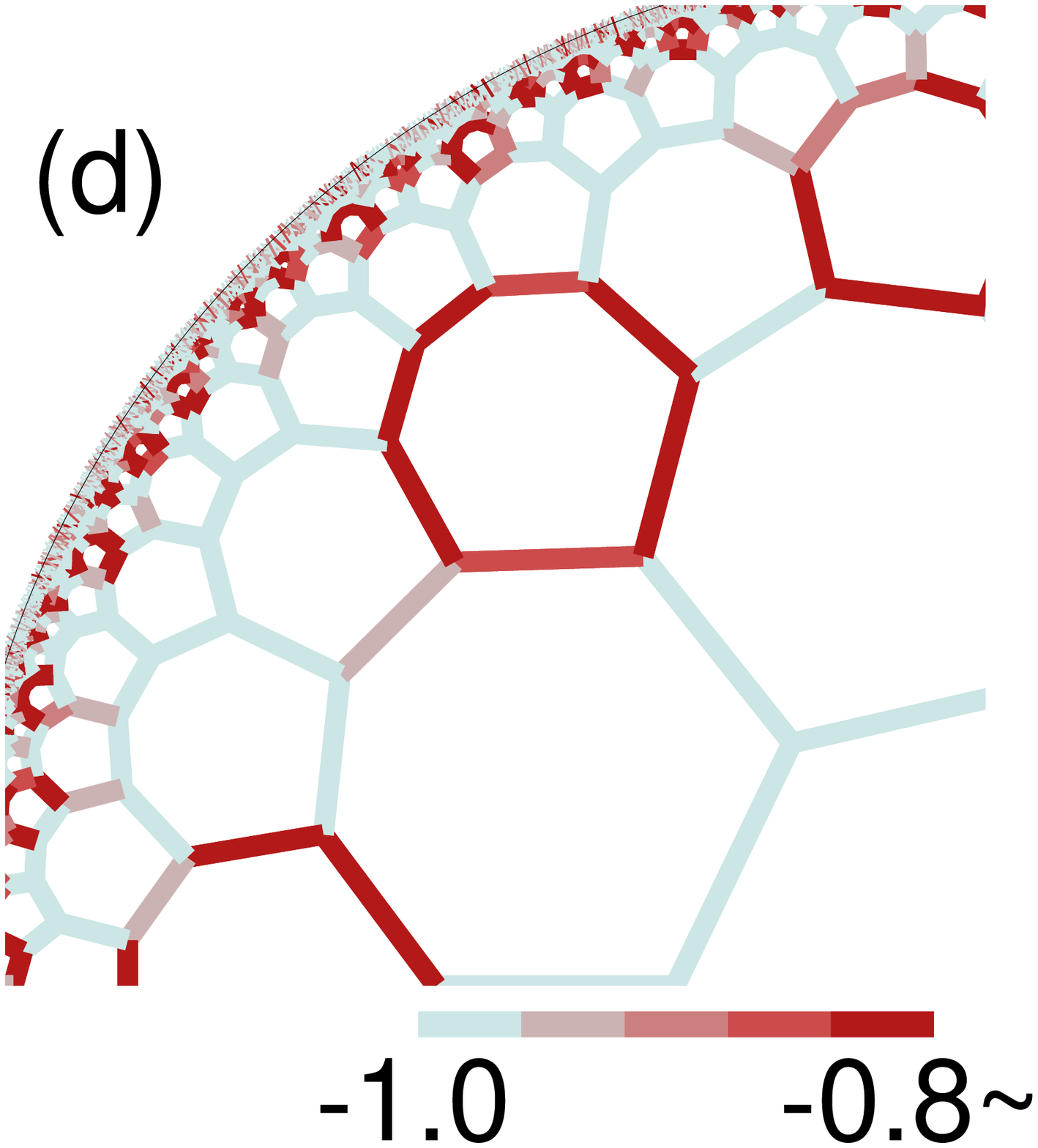}
\caption{
(Color online) Snapshots of spin configurations with $l=7$,
where a darker line represents a higher-energy bond.
(a) A high-temperature regime with $T=1.00$ and (b) a closer look at the
surface.
(c)
A low-temperature regime with $T\approx0.06$ and 
(d)
its magnified image, where clustering structures are clearly visible.
The bond energy and temperature $T$ are in units of $J$ and $J/k_B$,
respectively.
}
\label{fig:config}
\end{figure}

To equilibrate the system described by Eq.~(\ref{eq:h2}),
we employ MC simulations incorporated with
the
parallel
tempering (PT) method ~\cite{exchange}.
It is a numerical technique devised
to take averages efficiently over the state space.
In this method, MC calculations are carried out
simultaneously at different temperatures,
which significantly saves computational cost
for equilibrating large-scale frustrated systems.
Suppose that we have spin configurations $\{\phi_i\}_{T_k}$ and
$\{\phi_i\}_{T_{k+1}}$ running at temperatures $T=T_k$ and $T_{k+1}$,
respectively. Without loss of generality, we may set $T_k > T_{k+1}$.
Performing the standard Metropolis algorithm on the configurations, we
regularly check their energies $E\left[\{\phi_i\}_{T_k}\right]$ and
$E\left[\{\phi_i\}_{T_{k+1}}\right]$ and
exchange these configurations with a probability of $\min \left[1,
e^{- \left(E\left[\{\phi_i\}_{T_k}\right] -
E\left[\{\phi_i\}_{T_{k+1}}\right] \right) \left(1/T_{k+1} - 1/T_k \right)}
\right]$.
That is, when
$\{\phi_i\}_{T_{k+1}}$ is trapped in a local energy minimum,
PT makes it probe a wider region of the state space
by passing it to a higher temperature, $T_k$.
If it finds
a state with a sufficiently low energy, the vicinity can be checked in more
detail by lowering its temperature again.

Figure~\ref{fig:config} plots the resulting spin configurations at high and
low temperatures.
In the former two figures, 
there appears no regularity in the distribution of high-energy bonds 
(denoted by bonds with darker tone).
On the contrary,
at low $T$, high-energy bonds are merged into localized regions forming clusters.
It implies that the formation of such high-energy clusters is
energetically favorable:
even though frustration cannot be
removed, they may be shifted and confined into small
regions, lowering the energy in other regions instead.
Furthermore, 
we have observed 
that the locations of such
high-energy clusters remain floating by thermal fluctuations.
Thereby the distribution of high- and low-energy bonds is not
unique making the ground state degenerate.
Indeed, we find that each high-energy cluster in Fig.~\ref{fig:config}
appears as a topological defect, i.e., a negative vortex
to reduce the free-energy cost induced by the curvature~\cite{defects}.
It is noteworthy that those plots show a similarity to the results in
Ref.~\cite{sausset2008}, which reports that defects of a glass-forming
liquid in a negatively curved space are concentrated in local regions at low
$T$.
The origin of negative vortices
differs inherently from thermal dissociation of a vortex-antivortex pair
on a hyperbolic surface
whose properties were considered in Ref.~\cite{belo}.
Still, we note that if the interaction between vortices becomes short ranged
due to curvature~\cite{belo}, this effect may be relevant in exhibiting
glassy features.

\begin{figure}[ttt]
\includegraphics[width=0.4\textwidth]{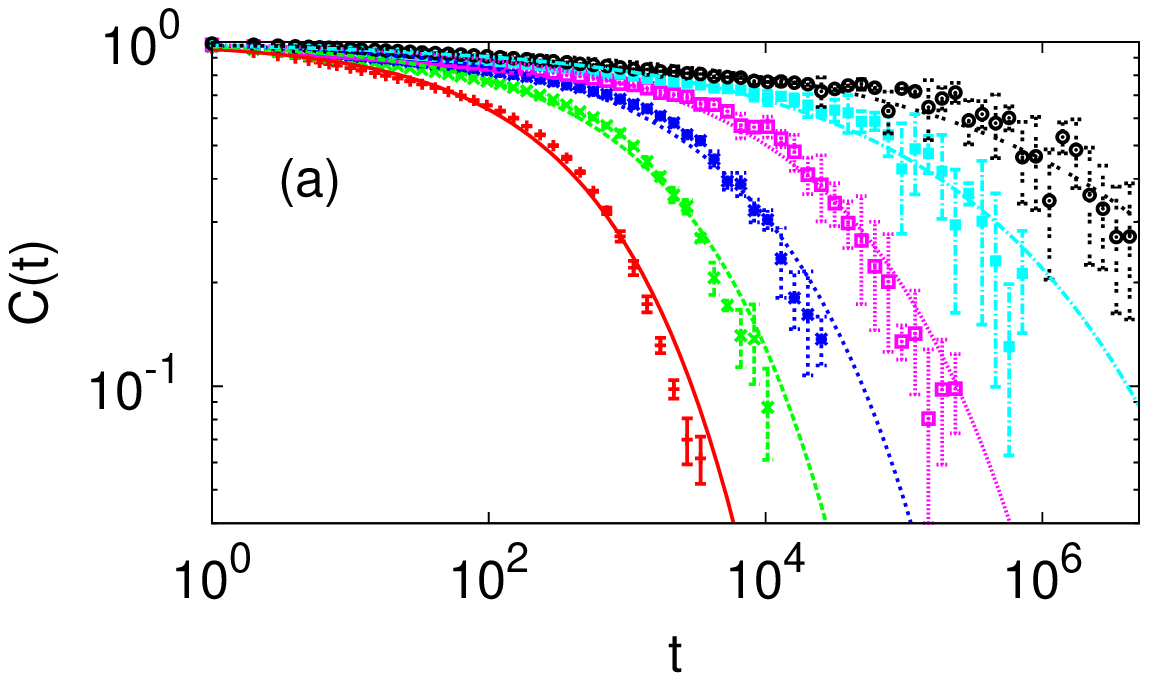}
\includegraphics[width=0.4\textwidth]{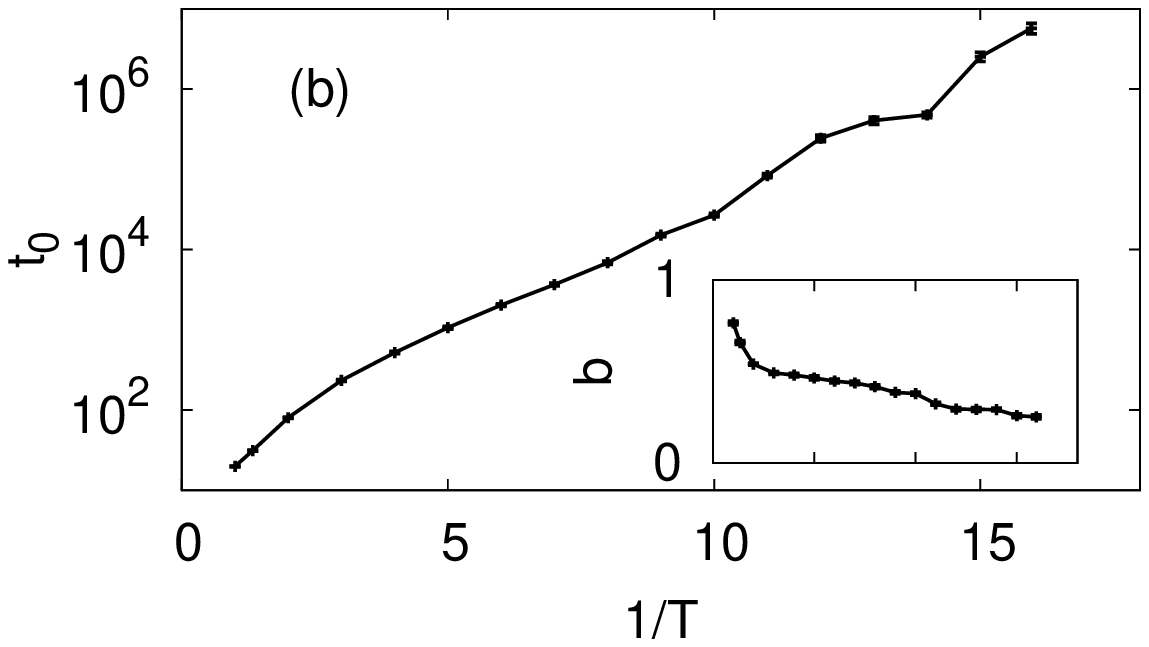}
\caption{(Color online) (a) Autocorrelations of spin configurations, given in Eq.~(\ref{eq:auto}),
together with fitting results with a stretched-exponential form. The results
are obtained for $T=0.250, 0.167, 0.125, 0.100, 0.083$, and $0.067$, from
bottom to top.
(b) Relaxation time $t_0$ as a function of inverse temperature from
the stretched-exponential fit. Inset: $b$ as a function of $1/T$ from the
same fit. It approaches unity at high $T$ leading to a simple
exponential form.
}
\label{fig:auto}
\end{figure}

The glassy behavior is quantified by the relaxation time $t_0$ for
the system to evolve from the configurations obtained in PT.
To this aim, we calculate autocorrelation
\begin{equation}
C(t) = \frac{1}{N \tau} \sum_{t'=1}^{\tau} \left| \sum_{j=1}^N
e^{i[\phi_j(t+t')-\phi_j(t')]} \right|,
\label{eq:auto}
\end{equation}
with the standard Metropolis algorithm for $\tau$
$= 10^4$
MC time steps.
We extract $t_0$ by fitting $C(t)$ with a stretched-exponential
function, $ C(t) \propto \exp [-(t/t_0)^b]$~\cite{phillips},
characterizing very slow relaxation in glasses.
Figures~\ref{fig:auto}(a) and \ref{fig:auto}(b) display
the measurements for $l=7$ and fitting results.
The relaxation behavior is described quite well by
the simple Arrhenius form, $t_0 \propto \exp(\Delta/T)$,
in which $\Delta$ indicates the activation energy.
It is noteworthy that the persistent linearity of $\log t_0$ with
respect to $1/T$
shown in Fig.~\ref{fig:auto}(b)
(i.e., $T$-independent activation energy $\Delta$)
is commonly observed in supercooled liquids such as $v$-{\rm SiO}$_2$
and $v$-{\rm GeO}$_2$,
in which spatiotemporal fluctuation of local configuration leads to critical
slowing down~\cite{deb}.

\begin{figure}[ttt]
\includegraphics[width=0.4\textwidth]{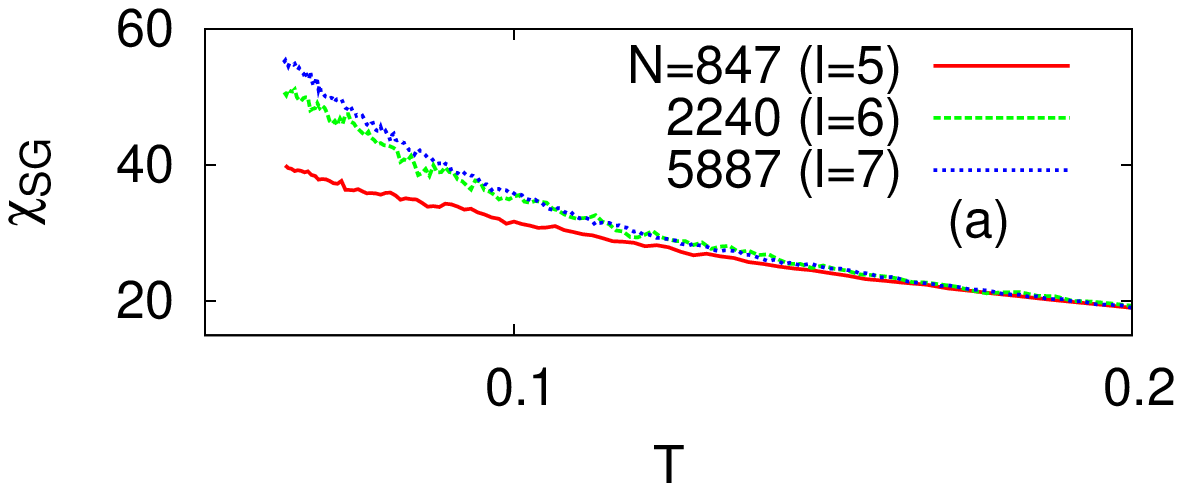}
\includegraphics[width=0.4\textwidth]{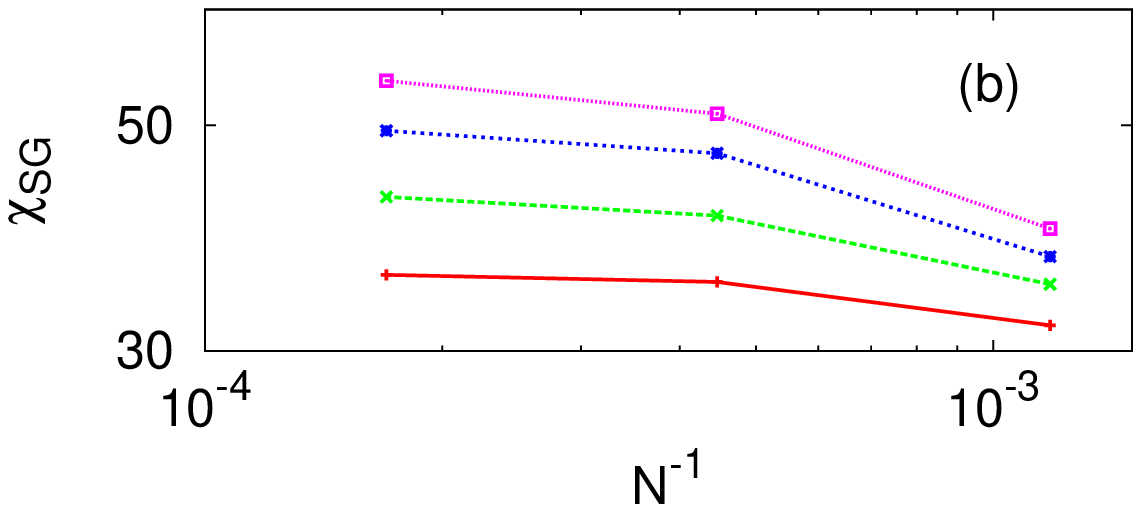}
\caption{(Color online) (a) Spin-glass susceptibility $\chi_{SG}$ as
a function of $T$. (b) $\chi_{SG}$ plotted against $N^{-1}$ at
$T=0.100, 0.083, 0.071$, and $0.063$, from bottom to top.
It does not exhibit a power law with respect to $N$ for any finite $T$.
}
\label{fig:sus}
\end{figure}

The presence of a glass transition as well as transition temperature
can be probed by observing the divergence of the spin-glass susceptibility
designated by $\chi_{SG}$.
In performing PT to calculate $\chi_{SG}$,
we prepare two replicas $\mu$ and $\nu$ for each $T$ and $N$.
Then we obtain the spin-glass susceptibility 
$\chi_{SG} = 
N
S(\tau;t_0')/\tau$, where
\begin{equation}
S(\tau;t_0') \equiv \sum_{t'=1}^{\tau} \left|
\frac1N
\sum_{j=1}^{N} e^{i[\phi_j^\mu(t_0'+t')-\phi_j^\nu(t_0'+t')]} \right|^2,
\end{equation}
with the measurement time $\tau$ after some equilibration time
$t_0'$~\cite{choi}. 
Figure~\ref{fig:sus}(a) gives the $T$ dependence of $\chi_{SG}(T,N)$ with
varying $N$'s. Our simulation code can properly handle the numerical
precision up to $l=7$, and the cases for $l<5$ are excluded due to
finite-size effects.
We see that $\chi_{SG}$ begins to increase at low $T$.
In order to evaluate the glass transition temperature $T_g$,
$\chi_{SG}$ is plotted against $N$ on a log-log scale in
Fig.~\ref{fig:sus}(b).
Strikingly, all the results of $\chi_{SG}$
give no sign of attaining a power law of $N$ at any finite $T$ we have tried.
This implies that a true singularity lies only at $T=0$.
We thus conclude that $T_g=0$; i.e.,
this system undergoes a zero-temperature glass transition.
More detailed finite-size analysis remains to be done largely due to
lack of available system sizes. One interesting point is that
a hyperbolic lattice does not possess scale invariance as it has its own
length scale, i.e., the radius of curvature~\cite{mode},
which may modify the nature of the singularity.

In summary, we have demonstrated an apparent zero-temperature orientational glass transition
in the $XY$ spin model on a negatively curved surface.
MC simulations revealed that
lowering $T$ makes high-energy bonds gather to form clusters,
which remain floating with very long relaxation times. 
The long lifetime fluctuation in cluster distribution follows
an Arrhenius-type relaxation at low $T$
and the singularity of spin-glass susceptibility is expected to
arise only at $T_g = 0$.
These observations are consequences of geometry of the surface,
where curvature-induced frustration in local (and thus global) spin configurations
yields ground-state degeneracy.

\acknowledgments
We are grateful to K. Nemoto for giving us helpful comments.
S.K.B. acknowledges the support from the Swedish Research Council 
under Grant No. 621-2002-4135
and B.J.K. was supported by the Korea Research Foundation Grant
under Grant No. KRF-2007-313-C00282.
H.S. is thankful for the support by
a Grant-in-Aid for Scientific Research from Japan Society
for the Promotion of Science (Grant No.~19360042).

\end{document}